\documentclass[12pt]{article}
\usepackage{graphicx}
\usepackage{hyperref}
\usepackage{lineno}
\usepackage{geometry}
\usepackage{amsmath}
\title{Exploring new physics in the dark sector at CMS \\~\\
\large Presented at the 32nd International Symposium on\\
Lepton Photon Interactions at High Energies,\\
Madison, Wisconsin, USA}
\author{Kai Hong Law \\ Imperial College London \\ \textit{On behalf of the CMS collaboration}}
\date{25-29 August 2025}

\begin{document}
\maketitle

\begin{abstract}
A selection of new results from the CMS experiment is presented. These results focus on searches for dark-sector particles using Run 2 or Run 3 data. Dedicated data streams were utilised to explore the low-mass parameter space. Machine learning techniques were employed to discriminate between signal and background.
\end{abstract}

\newpage

\section{Introduction}

One of the biggest open questions in the Standard Model (SM) of particle physics concerns the nature of dark matter. Hidden valley models are a class of models beyond the SM that extend the SM gauge group by a new, non-Abelian gauge group, which describes a new dark sector \cite{Strassler:2006im, HUR2011262}. Dark matter candidates are uncharged under the SM gauge group, however, there are mediator particles in these models that connect the dark sector to the SM. Hidden valley models can produce dark hadrons with high multiplicity, known as dark showers.

This report will focus on three new results from the CMS experiment \cite{CMS:2008xjf} which search for dark showers through displaced muons or lepton-enriched semi-visible jet signatures. Two of the searches use special data streams that were recorded during the Run 2 or Run 3 data taking period, which aimed at increasing the physics acceptance. The remaining search uses the full Run 2 data set.

\section{New Physics Results}

\subsection{Search for low-mass hidden-valley dark showers with non-prompt muon pairs}
\label{Parking-analysis}

This search is performed with a data set collected in 2018 using a ``data parking'' strategy \cite{CMS:2024zhe,Bainbridge:2730111}, in which the thresholds used by the trigger algorithms are lowered compared to standard data sets. This is enabled by delayed offline event reconstruction, which overcomes the limitation in computing resources to perform prompt reconstruction.  

Event-level Boosted Decision Trees (BDTs) are used to perform background rejection. Variables from muons, secondary vertices (produced from all types of tracks), and muon secondary vertices (produced from muon tracks) are used as input variables. A selection is applied on the BDT score after basic analysis selection criteria, and the BDT selection alone achieves $\sim$ $10^{4}$ background rejection with higher than $\sim30\%$ signal efficiency. 

Three broad classes of dark-sector models are studied: the vector portal model with a long-lived dark vector meson (denoted by $\tilde{\omega}$) \cite{Knapen:2021eip}; a model with a long-lived dark photon (denoted by $\textrm{A}'$), Scenario A \cite{Born:2023vll}; and a model with a long-lived dark pion (denoted by $\pi_{3}$), Scenario B1 \cite{Born:2023vll}. The production of dark showers from decays of the SM Higgs boson is considered.

\begin{figure}[h!] 
\centering
\includegraphics[width=0.45\textwidth]{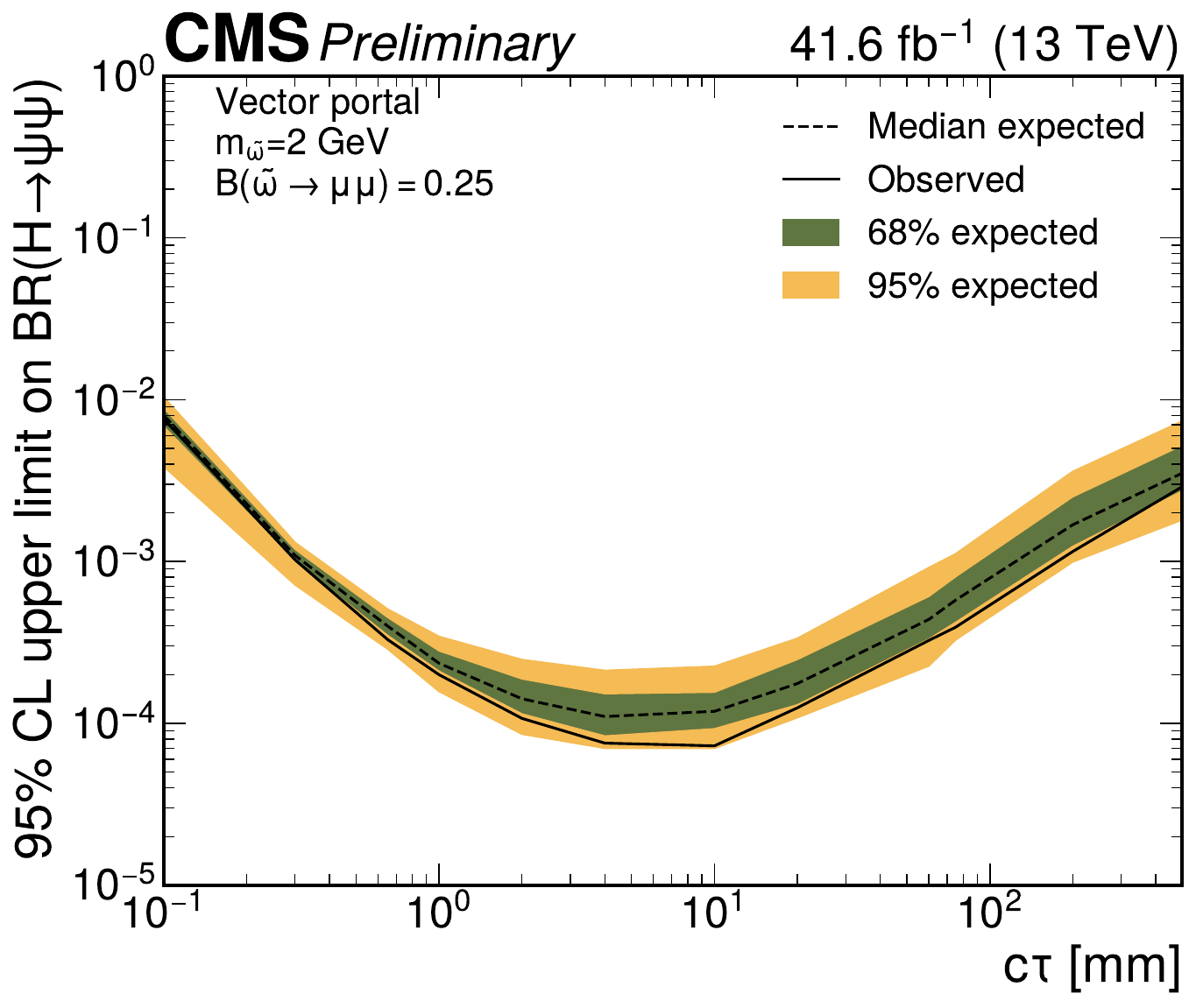}
\includegraphics[width=0.46\textwidth]{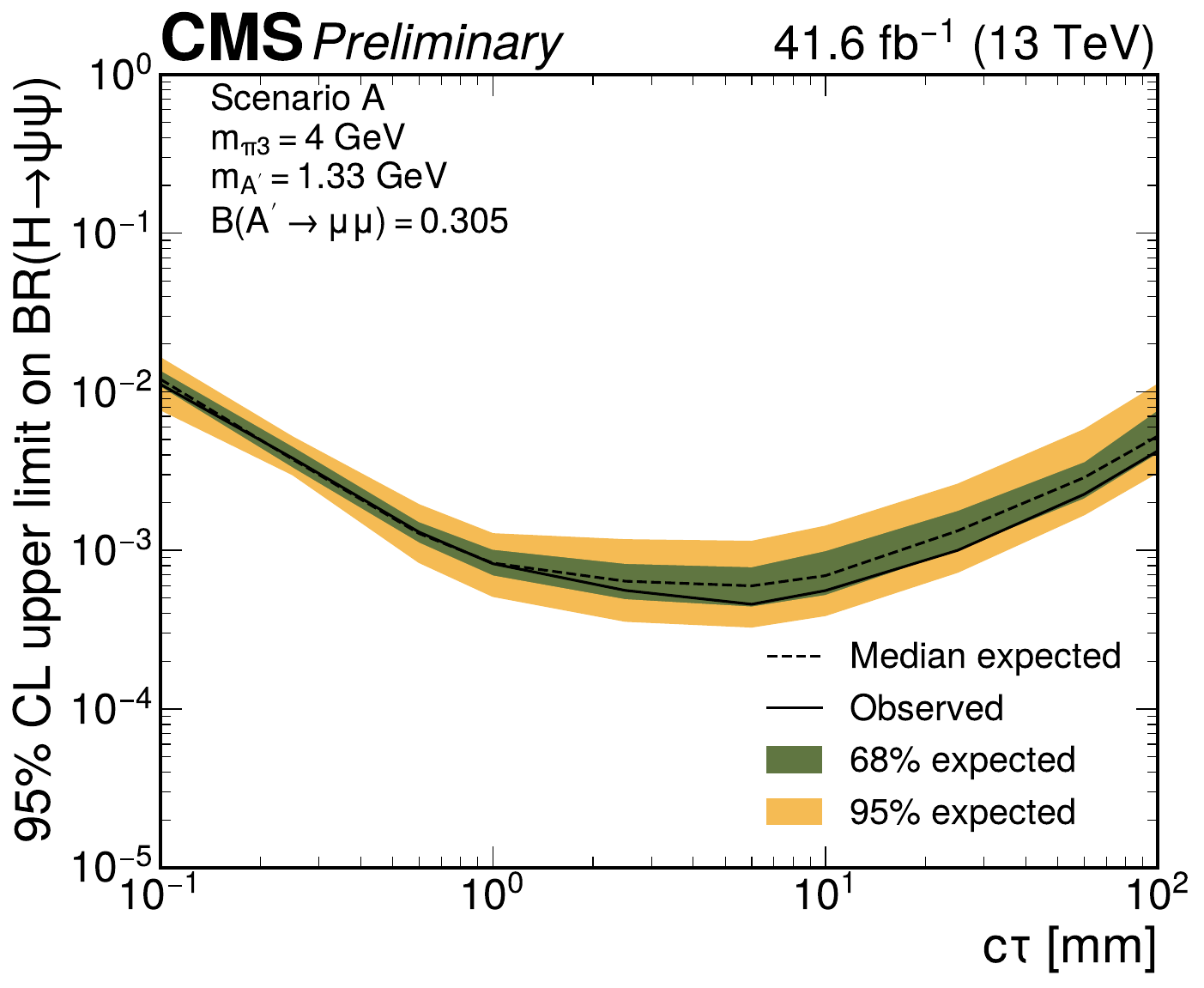}
\includegraphics[width=0.45\textwidth]{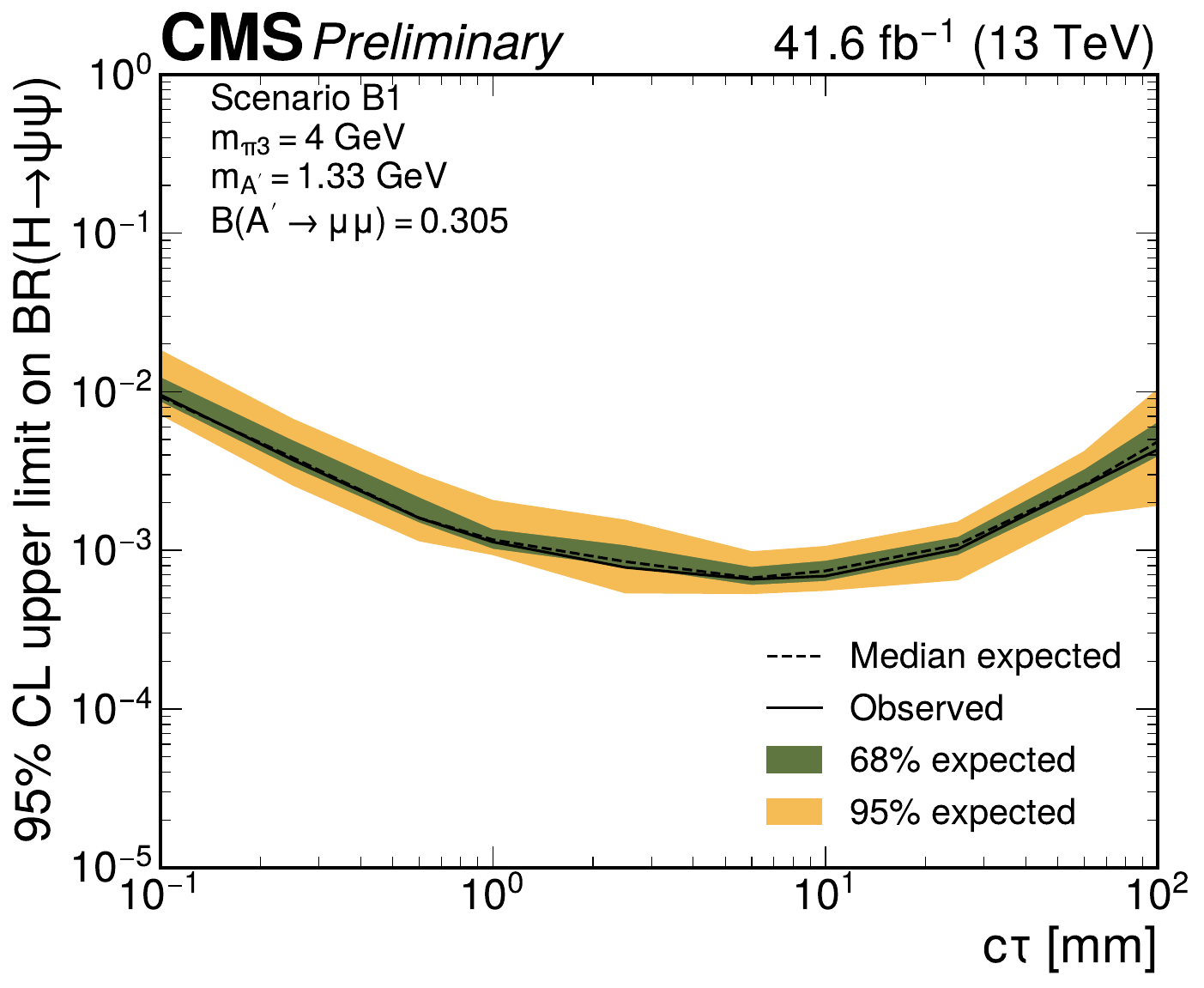}
\caption{Upper limits at $95\%$ CL on the branching fraction $\mathcal{B}\left(H\rightarrow\psi\overline{\psi}\right)$ for the vector portal as a function of the dark vector meson $c\tau$ (upper left), for Scenario A as a function of the dark photon $c\tau$ (upper right), and for Scenario B1 as a function of the dark pion $c\tau$ (lower). The limits are shown for an example mass hypothesis for each model \cite{CMS-PAS-EXO-24-008}.}
\label{parking_limits_vector_scenarioA_scenarioB1}
\end{figure}

Parametric fits are performed using the dimuon invariant mass distributions for different signal mass and lifetime hypotheses for each of the signal models. No significant excess beyond the SM expectation is observed, and $95\%$ confidence level upper limits are set on the branching fraction of the Higgs boson decaying into dark partons, $\mathcal{B}\left(H\rightarrow\psi\overline{\psi}\right)$. Figure \ref{parking_limits_vector_scenarioA_scenarioB1} shows the upper limits for the vector portal as a function of the mean proper decay length ($c\tau$) of $\tilde{\omega}$ meson, for Scenario A as a function of the $c\tau$ of $\textrm{A}'$, and for Scenario B1 as a function of the $c\tau$ of $\pi_{3}$ meson, respectively. The limits are shown for an example mass hypothesis for each model. Stringent limits are imposed on the branching fraction as low as $10^{-4}$. Good sensitivity is achieved for the Scenario B1 model, which is a non-pointing scenario featuring dimuon vertices that do not point back to the beam axis. The most stringent limits to date are imposed on the vector portal model with dark $\tilde{\omega}$ mesons of mean proper decay length below $500~\mathrm{mm}$ and masses between 0.3 and $20~\mathrm{GeV}$.

\subsection{Search for long-lived particles decaying into muons using the CMS scouting data sets}

This search is performed with data sets that were collected with a dedicated dimuon trigger stream (referred to as scouting \cite{CMS:2024zhe}) in 2022 and 2023. The triggers have high rate owing to less stringent requirements, and only a reduced amount of information is retained compared to standard data sets.  

The scouting data stream was improved in Run 3 compared to Run 2, providing access to new parameter space and increasing the amount of information stored in the events. The main improvement originates from the removal of the pixel hit multiplicity requirement for muons, which allows muon secondary vertices with $l_{xy}>11~\textrm{cm}$ to be accessed for the first time using scouting data \cite{CMS:2024zhe}. 

\begin{figure}[h!] 
\centering
\includegraphics[width=0.45\textwidth]{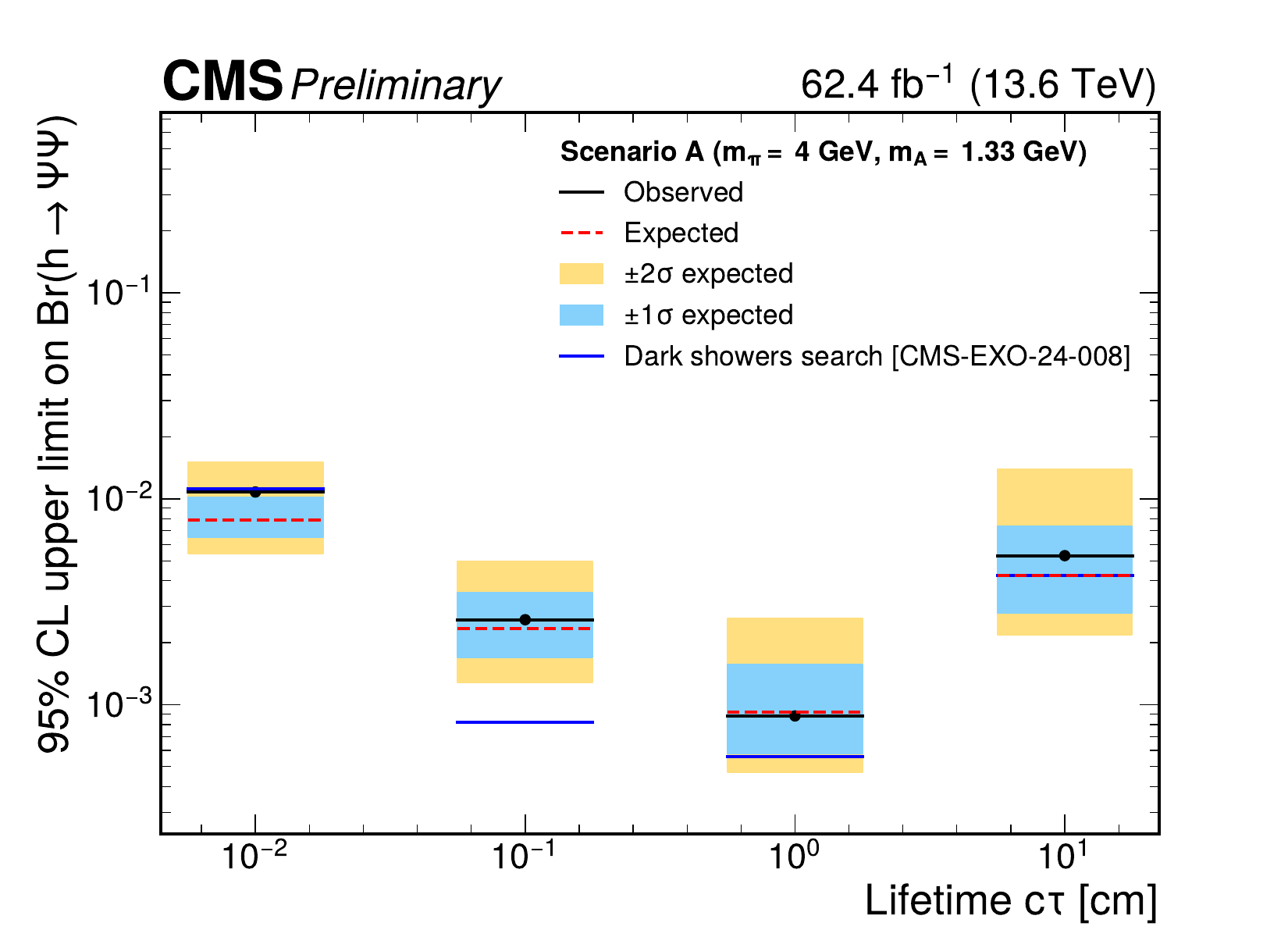}
\includegraphics[width=0.46\textwidth]{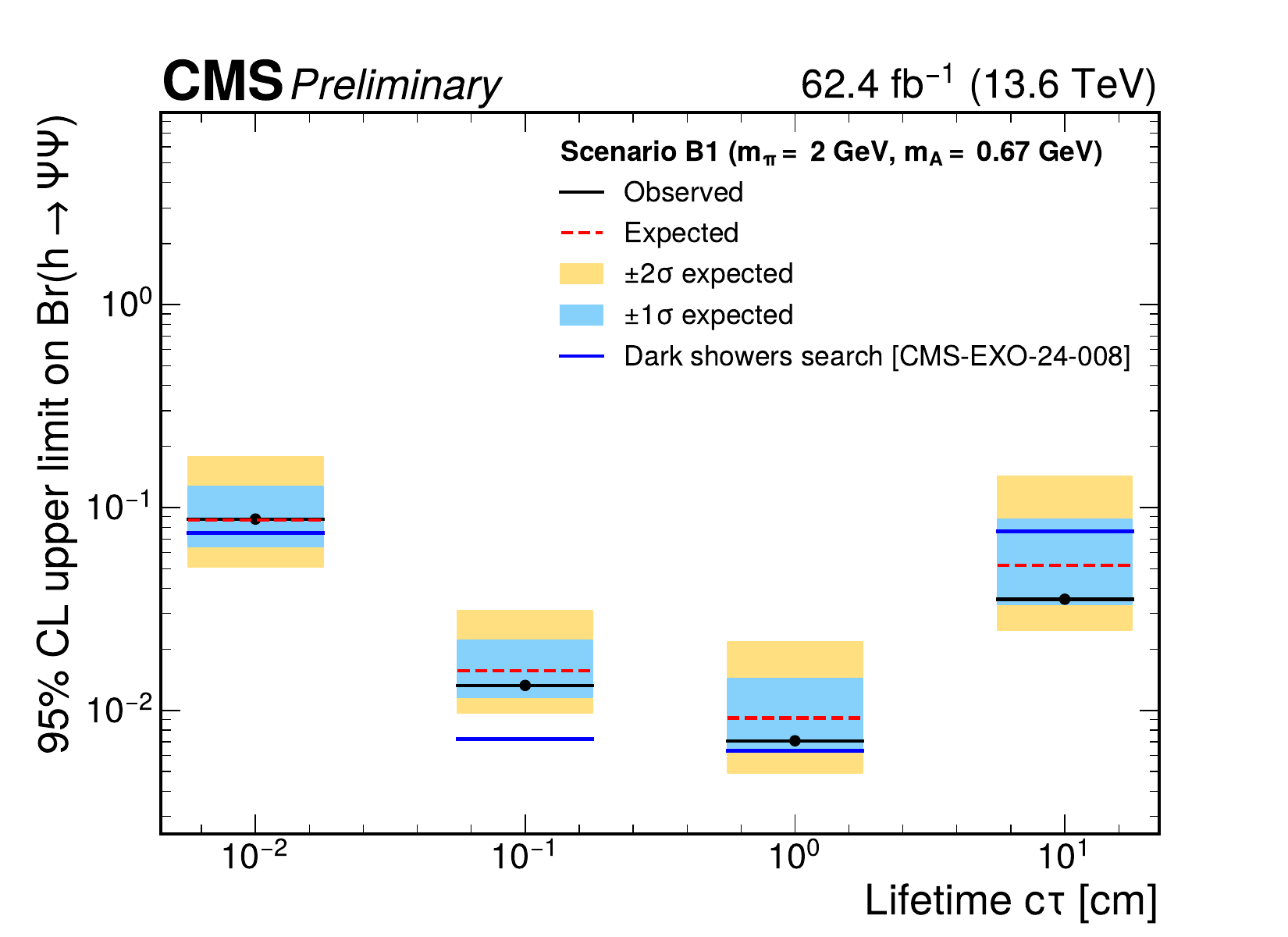}
\caption{Upper limits at $95\%$ CL on the branching fraction $\mathcal{B}\left(H\rightarrow\psi\overline{\psi}\right)$ for Scenario A as a function of the dark photon $c\tau$ (left), and for Scenario B1 as a function of the dark pion $c\tau$ (right). The limits are shown for an example mass hypothesis for each model \cite{CMS-PAS-EXO-24-016}. The dark blue solid line represents the observed upper limits set by Ref. \cite{CMS-PAS-EXO-24-008}.}
\label{limits_scenarioA_scenarioB1_comparison_with_scouting}
\end{figure}

The hidden-valley models Scenario A and Scenario B1 are considered in this analysis. The search is performed using parametric fits of the dimuon mass distributions for different signal mass and lifetime hypotheses. Figure \ref{limits_scenarioA_scenarioB1_comparison_with_scouting} shows the limits obtained for representative dark photon $c\tau$ hypotheses in Scenario A (left) and for representative dark pion $c\tau$ hypotheses in Scenario B1 (right). The limits are shown for an example mass hypothesis for each model. The corresponding observed limits from Ref. \cite{CMS-PAS-EXO-24-008} (the search presented in Section \ref{Parking-analysis}) are also shown in Fig. \ref{limits_scenarioA_scenarioB1_comparison_with_scouting}. While Ref. \cite{CMS-PAS-EXO-24-008} achieves a stronger sensitivity in general due to the use of machine learning techniques targeted at the signal models, this search provides complementary sensitivity for longer proper decay lengths due to the improvement in the scouting muon reconstruction at higher displacements.

\subsection	{Search for the resonant production of lepton-enriched semivisible jets}
Dark showers can produce visible jet-like final states produced by dark hadrons decaying into SM quarks. In the case where a fraction of the dark hadrons remain stable, a distinctive signature involving multijet plus missing transverse momentum ($p_{\textrm{T}}^{\textrm{miss}}$) is formed, where the $p_{\textrm{T}}^{\textrm{miss}}$ is aligned with one of the jets. This is referred to as the semivisible jet (SVJ) signature \cite{Cohen:2015toa}. This search is the first search for the resonant production of SVJs that contain leptons, which are known as lepton-enriched semivisible jets (SVJ$\ell$s) \cite{Cazzaniga:2022hxl}. 

 The process of interest is the resonant production of dark quarks from a $\textrm{Z}'$ boson. There are several parameters in the signal model, including: the $\textrm{Z}'$ mass, $m_{\textrm{Z}'}$; the dark hadron mass, $m_{\textrm{dark}}$; and the fraction of stable dark hadrons produced in the dark hadronisation process, $r_{\textrm{inv}}$. 

The search is performed using data sets collected in 2016-2018. A graph neural network jet identification algorithm (LundNet) \cite{Dreyer:2020brq} is utilised to discriminate between signal and background. A deep neural network is employed to extract uncorrelated discriminating features for estimating the background in the signal region using an ABCD method \cite{CMS:2025cvw}. This is the first time such techniques have been extended and employed for a resonant search.

\begin{figure}[h!] 
\centering
\includegraphics[width=0.6\textwidth]{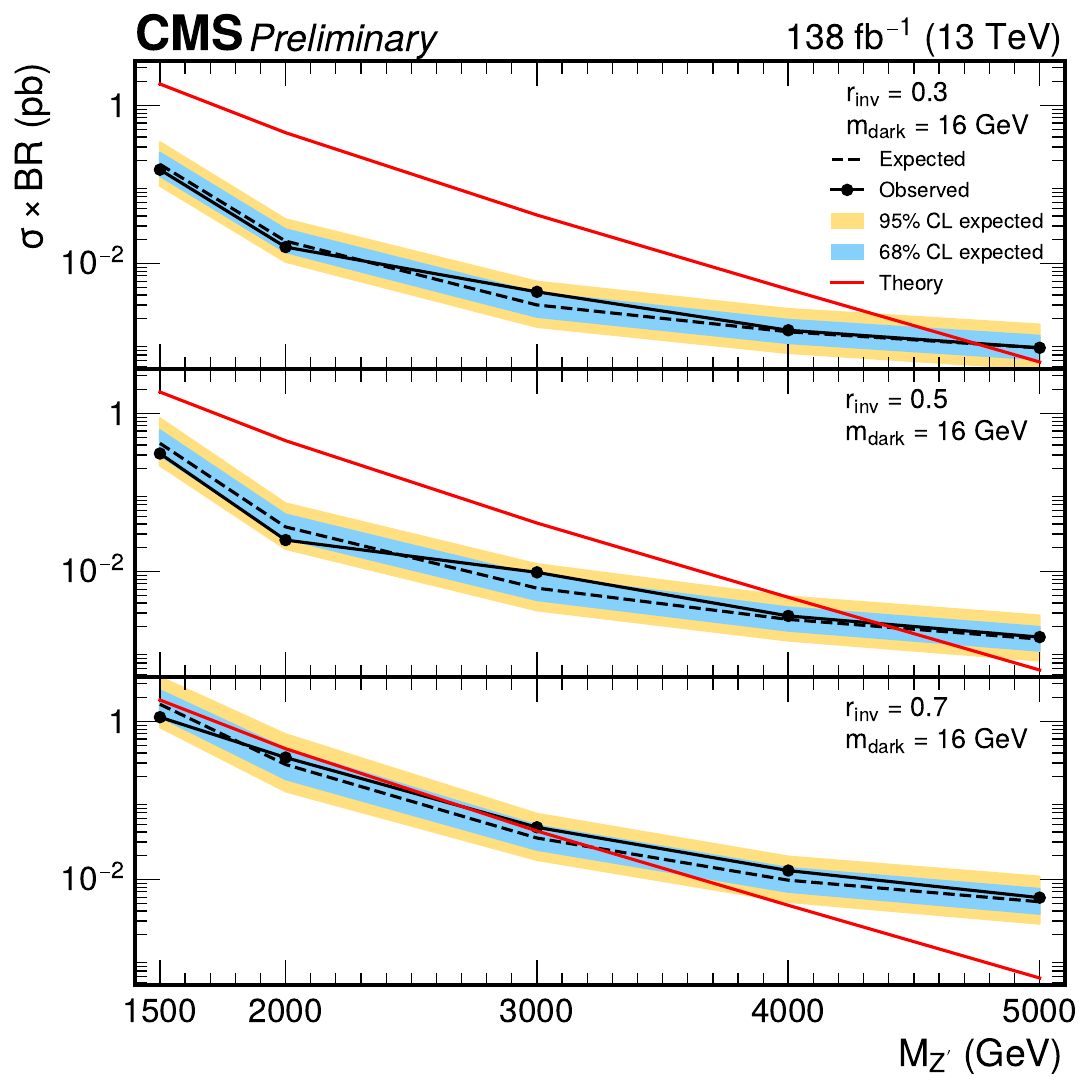}
\caption{Upper limits at $95\%$ CL on $\sigma_{\textrm{Z}'}\mathcal{B}_{\textrm{dark}}$ for the SVJ$\ell$ model as a function of $m_{\textrm{Z}'}$ for $r_{\textrm{inv}}=0.3, 0.5, 0.7$ and $m_{\textrm{dark}}=16~\textrm{GeV}$. The red solid line represents the nominal $\textrm{Z}'$ cross section \cite{CMS-PAS-EXO-24-029}.} 
\label{limits_SVJl}
\end{figure}

A binned fit is performed using the dijet transverse mass. No significant excess is observed above the expected background, and $95\%$ CL upper limits are set on the effective cross section $\sigma_{\textrm{Z}'}\mathcal{B}_{\textrm{dark}}$, which is the product of the cross section for $\textrm{Z}'$ boson production and the $\textrm{Z}'$ branching fraction to dark quarks. The limits are shown in Fig. \ref{limits_SVJl} as a function of $m_{\textrm{Z}'}$ for different values of $r_{\textrm{inv}}$ and $m_{\textrm{dark}}=16~\textrm{GeV}$. Models are excluded if the observed limit is below the theoretical prediction for $\sigma_{\textrm{Z}'}\mathcal{B}_{\textrm{dark}}$. The first mass exclusion limits on the SVJ$\ell$ signature are set for $m_{\textrm{Z}'}$ masses from $1.5~\textrm{TeV}$ up to $4.7~\textrm{TeV}$.

\section{Conclusion}
The CMS experiment has made multiple new contributions to exploring the dark sector, searching for challenging signatures. Dedicated data streams provide opportunities for exploring new parameter space. The Run 2 parking and the Run 3 scouting data sets were utilised to search for dark showers, and the analyses imposed complementary limits on three dark shower benchmark models. Furthermore, a first search for the SVJ$\ell$ signature has been performed using the full Run 2 data set. 

\bibliographystyle{cms_unsrt}
\bibliography{main}

\end{document}